\documentstyle{europhys}

\def\And{{\rm and\ }}

\def\stars{\bigskip\centerline{***}\medskip}

\newif\ifboo \boofalse

\def\Review#1{\boofalse{\it #1},}
\def\Name#1{{\sc #1},}
\def\Vol#1{\ifboo Vol. {\bf #1}\else{\bf #1}\fi}
\def\Year#1{\ifboo #1\else(#1)\fi}

\def\Page#1{\ifboo {\rm p. #1}\else{\rm #1}\fi}

\begin{document}
%
%
%
\euro{?}{?}{?}{1999}
\Date{1999}
\shorttitle{W. WERNSDORFER {\it et al.} NUCLEAR SPIN DRIVEN ETC.}
\title{Nuclear spin driven resonant tunnelling of magnetisation in Mn$_{12}$ acetate}
\author{W. Wernsdorfer\inst{1}, R. Sessoli\inst{2} \And D. Gatteschi\inst{2}}
\institute{
     \inst{1} Lab. de Mag. Louis N\'eel $-$ CNRS, BP166, 38042 Grenoble, France\\
     \inst{2} Dept. of Chemistry, Univ. of Florence, 
          Via Maragliano 77, 50144 Firenze, Italy}
\rec{}{}
\pacs{
\Pacs{75}{45$+$j}{Macroscopic quantum phenomena in magnetic systems}
\Pacs{75}{50 Tt}{Fine-particle systems}
\Pacs{75}{60 Ej }{Magnetisation curves, hysteresis, Barkhausen and related effects}
      }
\maketitle
\begin{abstract}
Current theories still fail to give a satisfactory explanation of the observed quantum 
phenomena in the relaxation of the magnetisation of the molecular cluster Mn$_{12}$ 
acetate. In the very low temperature regime, Prokof'ev and Stamp recently proposed that 
slowly changing dipolar fields and rapidly fluctuating hyperfine fields play a major role in 
the tunnelling process. By means of a faster relaxing minor species of Mn$_{12}$ac and 
a new experimental 'hole digging' method, we measured the intrinsic line width 
broadening due to local fluctuating fields, and found strong evidence for the influence of 
nuclear spins on resonance tunnelling at very low temperatures (0.04 $-$ 0.3~K). 
At higher temperature (1.5 $-$ 4~K), we observed a homogeneous line width 
broadening of the resonance transitions being in agreement with a recent calculation of 
Leuenberger and Loss.
\end{abstract}
%
%
%
%
%
Observation of mesoscopic quantum phenomena in magnetism 
has remained a challenging problem. The first striking demonstration of 
quantum tunneling and quantum phase interference was found on 
Mn$_{12}$ acetate and Fe$_8$, molecular clusters having a spin ground state $S = 10$ 
\cite{Sessoli93, Friedman96, Sangregorio97, WW_RS99}. Several models and theories have been proposed to explain 
in detail the experimental results, published during the last five years by several authors 
\cite{theories}, but there is not yet satisfactory agreement between theory and 
experiments concerning mainly the relaxation rate and the resonance line width 
\cite{Friedman98}. This letter is intended to report more accurate measurements which 
should help to find a satisfactory explanation of the observed quantum phenomena. 

Several authors have pointed out that in the Mn$_{12}$ carboxylate family
different isomeric forms give rise to different relaxation rates \cite{Sun98}.
This has also been observed in Mn$_{12}$ acetate \cite{Aubin97}. 
We found that a minor species Mn$_{12}$ac(2) \cite{remark1}, randomly
distributed in the crystal of the major species Mn$_{12}$ac(1), exhibits a faster
relaxation rate which becomes temperature independent below 0.3 K. Even if this
second species has been only partially characterised \cite{remark1} we can exploit
it as a local probe providing unique information on the tunnelling process. We 
used a recently developed method \cite{Ohm98,Wernsdorfer99} for measuring the 
intrinsic line width broadening due to local fluctuating fields and found strong evidence 
for the influence of nuclear spins on resonance tunnelling. 

In the first part of this letter, we focus on the low temperature and low field limit 
which is particularly interesting because phonon-mediated relaxation is astronomically 
long and can therefore be neglected. In this limit, only the two lowest levels with quantum numbers
$M_z$~=$\pm$10 are involved. They are coupled by a tunnel matrix element $\Delta/2$ 
where $\Delta$ is the tunnel splitting which is estimated to be about 10$^{-10}$~K for 
Mn$_{12}$ \cite{Prok98}. In an ideal system, resonant tunnelling requires that the 
magnetic field (local to the giant spin) is smaller than the field associated with the tunnel 
splitting $\Delta$ which means a field smaller than 10$^{-9}$~T for the Mn$_{12}$ac 
clusters. This fact would make it very difficult to observe the tunnelling of isolated giant 
spins at constant applied field. This dilemma is solved by invoking the dynamics of dipolar interaction between 
molecules and nuclear spins \cite{Prok98}. The tunnelling scenario can be summarised as 
follows: the rapidly fluctuating hyperfine field brings molecules into resonance. The 
dipolar field of tunnelled spins can lift the degeneracy and remove from resonance a large 
number of neighbouring spins. However, a gradual adjustment of the dipolar fields 
across the sample (up to 0.03 T in Mn$_{12}$ac), caused by tunnelling relaxation, 
brings other molecules into resonance and allows continuous relaxation. Therefore, one 
expects a fast relaxation at short times, and slow logarithmic relaxation at long times.

\begin{figure}[t]
\centerline{\epsfxsize=6 cm \epsfbox{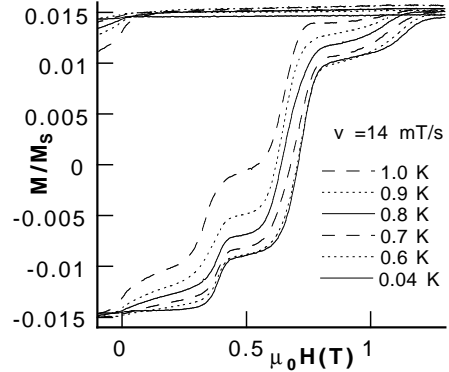}}
\caption{Magnetic hysteresis curves of the minor species Mn$_{12}$ac(2) \cite{remark1} at several temperature. Before the measurement, the major species Mn$_{12}$ac(1) was 
demagnetised \cite{remark2}. Resonant tunnelling is evidenced by four steps. Notice that the loops are about temperature independent below 0.6 K. $M_S$ is the magnetisation of 
saturation of the whole crystal.}
\label{fig1}
\end{figure}

Recently, we developed a method \cite{Ohm98,Wernsdorfer99} for measuring 
the intrinsic line width broadening due to local fluctuating fields of the nuclear spins. It is 
based on the general idea that the short time relaxation rate is directly connected to the 
number of molecules which are in resonance at a given longitudinal applied field $H$. 
The Prokof'ev - Stamp theory \cite{Prok98} predicts that the magnetisation should 
relax at short times with a square-root time dependence:

\begin{equation}
	M(H,t)=M_{\rm in}+(M_{\rm eq}(H)-M_{\rm in})\sqrt{\Gamma_{\rm 
	sqrt}(H)t} \label{eq1}
\end{equation}

Here $M_{\rm in}$ is the initial magnetisation at time $t$~= 0 (i.e. after a 
rapid field change), and $M_{\rm eq}(H)$ is the equilibrium 
magnetisation. The rate function $\Gamma_{\rm sqrt}(H)$ is 
proportional to the normalised distribution $P(H)$ of molecules which are in resonance at 
the applied field H:

\begin{equation}
	\Gamma_{\rm sqrt}(H)\sim\frac{\Delta^2}{\hbar}P(H) 
	\label{eq2}
\end{equation}

where $\hbar$ is Planck's constant. Thus, the measurements of $\Gamma_{\rm 
sqrt}(H)$, as a function of $H$, should give direct access to the distribution $P(H)$. 
Our measuring procedure is as follows. Starting from a well defined magnetisation state 
\cite{remark2}, we apply a magnetic field $H$ in order to measure the short-time square 
root relaxation behaviour. By using eq. (1), we get the rate function $\Gamma_{\rm 
sqrt}(H)$ at the field $H$. Then, starting again from the same well defined magnetisation 
state, we measure $\Gamma_{\rm sqrt}(H)$ at another field $H$, yielding the field 
dependence of $\Gamma_{\rm sqrt}(H)$ which is proportional to the dipolar distribution 
$P(H)$ (eq. (2)).

\begin{figure}[t]
\centerline{\epsfxsize=6 cm \epsfbox{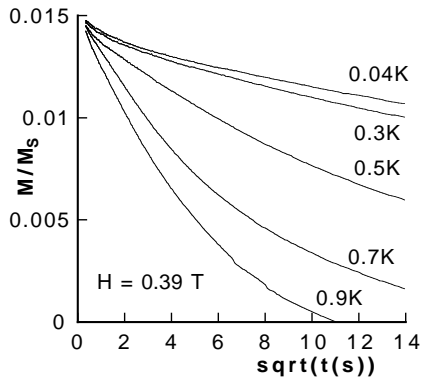}}
\caption{Typical relaxation measurements of the minor species Mn$_{12}$ac(2) \cite{remark1} measured at H = 0.39 T and for several temperatures. For each curve, the major species 
Mn$_{12}$ac(1) was demagnetised \cite{remark2} and Mn$_{12}$ac(2) was saturated in a field of -1.4 T. The $\Gamma_{\rm sqrt}$ relaxation rate was approximately found in the range from $0.014 > M/M_S > 0.01$, i.e. in the short time region. $M_S$ is the magnetisation of saturation of the whole crystal.}
\label{fig2}
\end{figure}

This technique can be used for following the time evolution of molecular states in 
the sample during a tunnelling relaxation \cite{Wernsdorfer99}. Starting from a well 
defined magnetisation state \cite{remark2}, and after applying a field $H_{dig}$, we let 
the sample relax for a time $t_{dig}$, called 'digging field and digging time', 
respectively. During the digging time, a small fraction of the molecular spins tunnel and 
reverse the direction of their magnetisation. Finally, we apply a field $H$ to measure the 
short time relaxation in order to get $\Gamma_{\rm sqrt}(H)$ (eq. (1)). The entire 
procedure is then repeated to probe the distribution at other fields $H$ yielding 
$\Gamma_{\rm sqrt}(H,H_{dig},t_{dig})$ which is proportional to the number of spins 
which are still free for tunnelling. With this procedure one obtains the distribution 
$P(H,H_{dig},t_{dig})$, which we call the 'tunnelling distribution'.

We used this new technique, which we call 'hole digging' method \cite{remark3}, 
for studying Fe$_8$ molecular clusters \cite{Wernsdorfer99} and found that tunnelling 
causes rapid transitions of molecules near $H_{dig}$, thereby "digging a hole" in 
$P(H,H_{dig},t_{dig})$ around $H_{dig}$, and also pushing other molecules away 
from resonance. The hole widens and moves with time, in a way depending on sample 
shape; the width dramatically depends on thermal annealing of the magnetisation of the 
sample. For small initial magnetisation \cite{remark2}, the hole width shows an intrinsic 
broadening which may be due to nuclear spins \cite{Wernsdorfer99}. 

\begin{figure}[t]
\centerline{\epsfxsize=6 cm \epsfbox{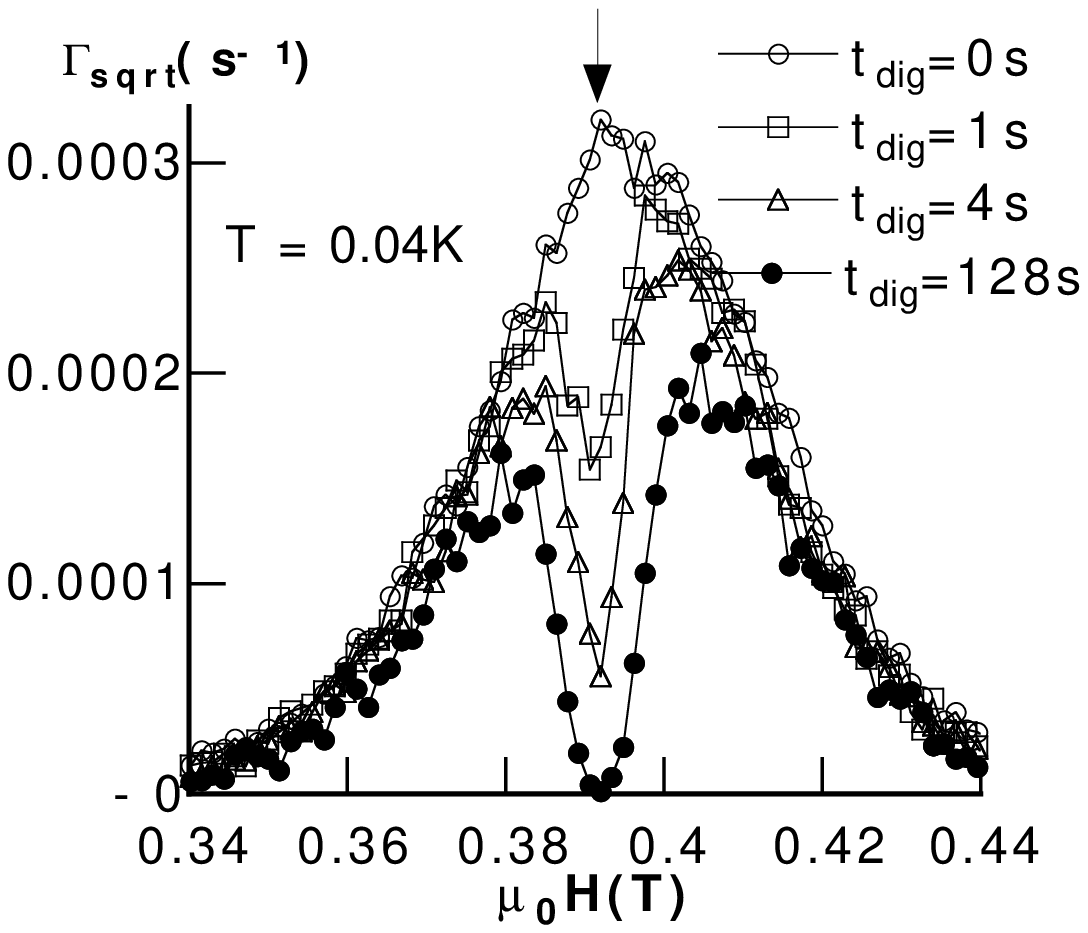}}
\centerline{\epsfxsize=6 cm \epsfbox{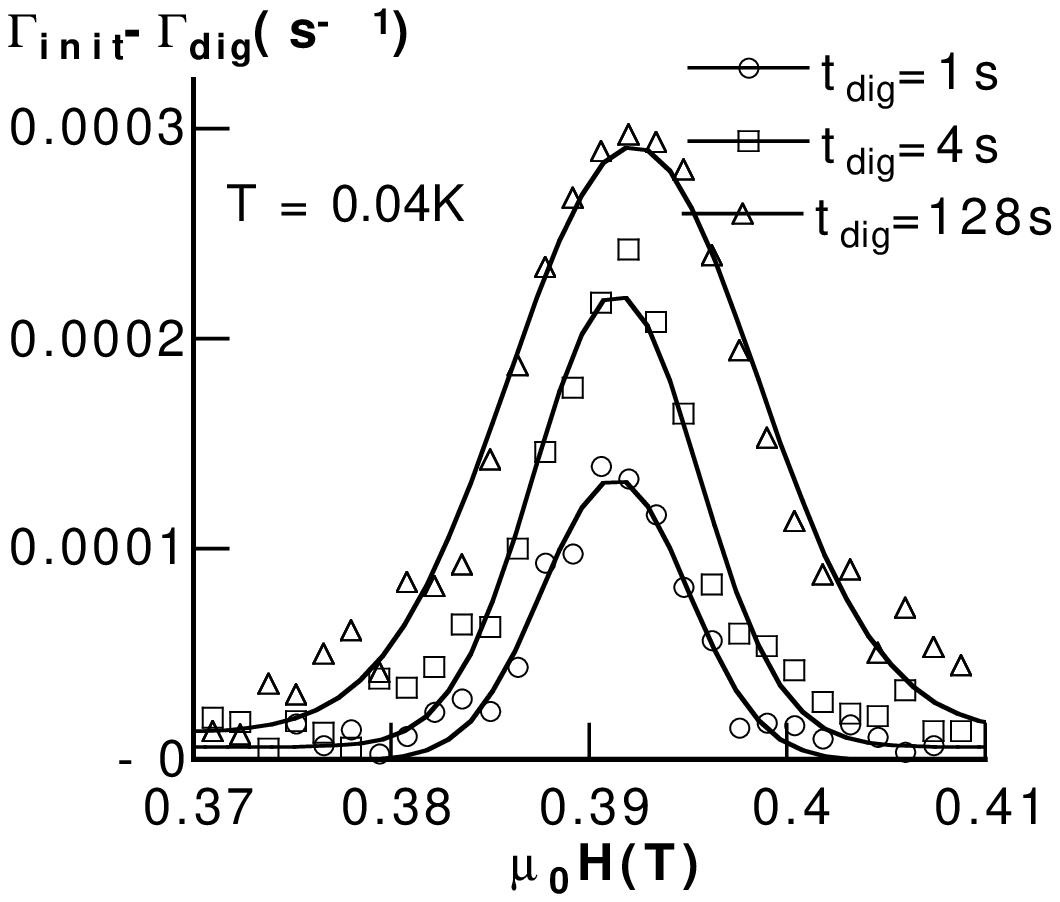}}
\caption{(a): Quantum hole digging in an initial dipolar distribution of the minor species 
Mn$_{12}$ac (2). (b): Difference between the initial dipolar distribution $\Gamma_{\rm init}$ and quantum tunnel distribution $\Gamma_{\rm dig}$ of Fig 3(a). The continuos lines are Gaussian functions fitted to the 
data yielding the hole line width $\sigma$.}
\label{fig3}
\end{figure}

For Mn$_{12}$ac(1) \cite{remark1}, T $<$ 1.5K and small applied fields, relaxation 
measurements are very time consuming because the pure quantum relaxation rate between 
$M_z$~=$\pm$10 is of the order of years or longer \cite{Wernsdorfer96}. However, we 
found that one can use the minor species Mn$_{12}$ac(2) \cite{remark1} which showed 
to have a much faster tunnelling rate. Furthermore, it has the advantage of being diluted 
over the entire crystals with a concentration of 1 to 2 percent, thus the internal dipolar 
fields hardly change during relaxation of Mn$_{12}$ac(2). As Mn$_{12}$ac(2) 
experiences the same environment (concerning mainly hyperfine fields) as the major 
species Mn$_{12}$ac(1), we propose to use Mn$_{12}$ac(2) as a local probe of any 
fluctuating field acting on the giant spins of Mn$_{12}$ac molecular clusters.

Below about 1.5 K, we found that the magnetisation of Mn$_{12}$ac(2) can be 
reversed in an applied field smaller than 2 T whereas that of Mn$_{12}$ac(1) hardly 
reverses because of the very small tunnelling rate. Fig. 1 presents a typical hysteresis loop 
measurements of Mn$_{12}$ac(2) which is almost temperature independent below 0.6 K 
\cite{remark4}. These loops are strongly field sweeping rate dependent and show 
quantum tunnelling resonances at about equidistant fields of $\Delta$H $\approx$ 0.39 T in comparison to 0.45 T for Mn$_{12}$ac(1). 

Fig. 2 presents typical relaxation measurements of the minor species 
Mn$_{12}$ac(2). For each curve, the major species Mn$_{12}$ac(1) was demagnetised 
\cite{remark2} and Mn$_{12}$ac(2) was saturated in a field of -1.4 T. Approximate square 
root relaxation was found in the range from $0.014 > M/M_S > 0.01$, where $M_S$ is the 
magnetisation of saturation of the entire crystal. The fact that the relaxation is not exactly 
proportional to $\sqrt{t}$ in the short time region, is irrelevant for the discussion of this letter \cite{remark5}.

Fig. 3a presents tunnelling distributions for Mn$_{12}$ac(2) for digging times 
between t$_0$ = 0 and 128 s. Note the depletion ("hole digging") around the digging field $H_{dig}$ = 0.39 T. This hole-digging arises because only spins in resonance can 
tunnel. Although the hole is narrow, it is still several orders of magnitude larger than the 
field associated with the tunnel splitting $\Delta$. The hole could be fitted to a Gaussian 
function yielding the line width $\sigma$ (see Fig. 3b) which we studied as a function of 
temperature and digging time (fig. 4). We defined an intrinsic line width $\sigma_0$ by a linear 
extrapolation of the curves to $t_{dig}$ = 0. For temperatures between 0.04 and 0.3K, $\sigma_0$ $\approx$ 12 mT. For T $>$ 0.3 K, $\sigma_0$ increase rapidly.

The physical origin of the line width $\sigma_0$ is tentatively assigned to the fluctuating 
hyperfine fields. A simple calculation of the random hyperfine field distribution was made 
by Hartmann-Boutron {\it et al.} \cite{Hartmann96}, who evaluated the maximum nuclear field 
operating on the lowest $M$ = 10 levels of Mn$_{12}$. Using the same approach it is 
possible to calculate the whole spectrum of hyperfine levels. Assuming  for the hyperfine 
coupling constants the values a(MnIII)= 6.9 mT and a(MnIV)= 8.5 mT, in agreement 
with currently accepted values for these ions \cite{Zheng96}, we calculate a Gaussian 
distribution of fields with a width of ca. 16 mT, in good agreement with the above 
reported experimental result. A detailed calculation of the random hyperfine field 
distribution can be found in Ref. \cite{Prok98}.

\begin{figure}[t]
\centerline{\epsfxsize=5.5 cm \epsfbox{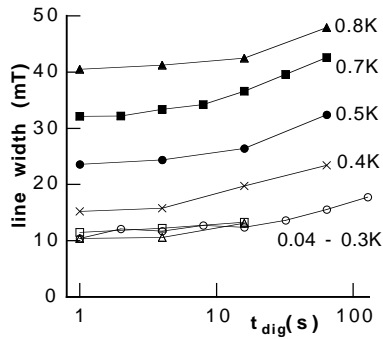}}
\caption{Hole line width $\sigma$, obtained by the measurements of quantum hole digging like in Fig. 3 (a) and (b), as a function of digging time for several temperatures. The intrinsic 
line width $\sigma_0$ is obtained by an linear extrapolation to $t_{dig}$ = 0 s.}
\label{fig4}
\end{figure}

We applied also the 'hole digging' method at temperatures between 1.5 and 4 
K. At these temperatures, the relaxation rates of the minor species are very fast and can 
therefore be neglected. As pointed out by several groups, the relaxation of the major 
species Mn$_{12}$ac(1) is non-exponential at temperatures below 4 K but, nevertheless, 
we approximately adjusted an exponential law for the short time relaxation regime (1 $-$ 
100s) in order to yield a relaxation rate $\Gamma$. We emphasise that the Prokof'ev 
Stamp theory \cite{Prok98} cannot be applied in the higher temperature regime because it 
neglects thermal activation to higher energy levels. However, the main idea of the 'hole 
digging' method should still hold, i.e. it should answer the question of whether the line 
width is homogeneously or inhomogeneously broadened. A typical result of the 'hole 
digging' experiment at 2 K is presented in Fig. 5 which shows that it is impossible to dig 
a hole in the $\Gamma(H)$ dependence suggesting that the line width is homogeneously 
broadened, as first suggested by Friedman {\it et al.} \cite{Friedman98}. This finding is also in good 
agreement with a recent calculation of Leuenberger and Loss \cite{Leuenberger99}, see also \cite{Fort98} 
which is based on thermally assisted spin tunnelling induced by quadratic anisotropy and 
weak transverse magnetic fields. Their model is minimal in the sense that it is sufficient to 
explain the measurements without including hyperfine fields. Indeed, our measurements 
show that the inhomogeneous hyperfine broadening of about 12 mT is small compared to 
the homogeneous broadening of about 30 mT (see fig. 5) which might be due to spin-
phonon coupling \cite{remark6}.

In conclusion, this letter is intended to report more accurate measurements which 
should help to find a satisfactory explanation of the observed quantum phenomena in 
molecular clusters. We used the recently developed 'hole digging' method for measuring 
the intrinsic line width broadening due to local fluctuating fields and found strong 
evidence for the influence of nuclear spins in the tunnel process at low temperature. At 
higher temperatures, spin-phonon coupling seems to dominate the resonant quantum 
tunnel transitions which leads to homogeneously broadened line widths.

\stars

	A. Caneschi is acknowledged for providing the Mn$_{12}$ac samples. We are 
deeply indebted to P. Stamp for many fruitful discussions. We thank B. Barbara, A. Benoit, E. Bonet Orozco, D. Mailly, and P. Pannetier for the help of constructing our micro-SQUID magnetometer.

\begin{figure}[t]
\centerline{\epsfxsize=5.5 cm \epsfbox{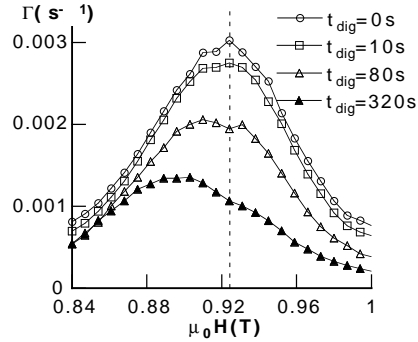}}
\caption{Quantum hole digging in an initial dipolar distribution of the major species 
Mn$_{12}$ac (1), at T = 2 K. The relaxation rate (found by an exponential short time fit) 
is plotted versus field $H$ after a digging relaxation at $H_{dig}$ = 0.925 T. After $t_{dig}$ = 10, 80 and 320 s, the reversed fraction of magnetisation is 0.042, 0.214, 0.573 of $M_S$, 
respectively. The shift of the maximum of $\Gamma(H)$ is due to the change of the 
internal dipolar fields.}
\label{fig5}
\end{figure}


\vskip-12pt
\begin{thebibliography}{99}
%
\bibitem{Sessoli93} \Name{Sessoli R., Gatteschi D., Caneschi A. \And Novak M.A.} \Review{Nature (London)} \Vol{365} \Year{1993} \Page{141}; \Name{Caneschi A. {\it et al.}} \Review{J. Magn. Magn. Mat.} \Vol{177-181} \Year{1998} \Page{1330}.

\bibitem{Friedman96} \Name{Friedman J. R., Sarachik M. P., Tejada J. \And Ziolo R.} \Review{Phys. Rev. Lett.} \Vol{76} \Year{1996} \Page{3830};
\Name{Thomas L. {\it et al.}} \Review{Nature (London)} \Vol{383} \Year{1996} \Page{145}.

\bibitem{Sangregorio97} \Name{Sangregorio C., Ohm T., Paulsen C, Sessoli R. \And Gatteschi D.} \Review{Phys. Rev. Lett.} \Vol{78} \Year{1997} \Page{4645}.

\bibitem{WW_RS99} \Name{Wernsdorfer W. \And Sessoli R.} \Review{Science} \Vol{284} \Year{1999} \Page{133}.

\bibitem{theories} \Name{Villain  J. {\it et al.} } \Review{Europhys. Lett.} \Vol{27} \Year{1994} \Page{159}; 
\Name{Hartmann-Boutron  F. {\it et al.}} \Review{Int. J. Mod. Phys. B} \Vol{10} \Year{1996} \Page{2577}; 
\Name{Garanin D. A. \And Chudnovsky E. M.} \Review{Phys. Rev. B} \Vol{56} \Year{1997} \Page{11102}; 
\Name{V.V. Dobrovitski and A.K. Zvezdin} \Review{Euro. Phys. Lett.} \Vol{38} \Year{1997} \Page{377}; 
\Name{Gunther L.} \Review{Euro. Phys. Lett.} \Vol{39} \Year{1997} \Page{1}; 
\Name{Fort  A. {\it et al.}} \Review{Phys. Rev.Lett.} \Vol{80} \Year{1998} \Page{612}; 
\Name{Luis E. {\it et al.}} \Review{Phys. Rev. B} \Vol{57} \Year{1998} \Page{505}; 
\Name{Garcia-Pablos D., Garia  N. \And De Raedt H.} \Review{Europhys. Lett.} \Vol{42} \Year{1998} \Page{473}; 
\Name{Garg A.} \Review{Phys. Rev. Lett.} \Vol{81} \Year{1998} \Page{1513}; 
see also the references given in ref. \cite{Friedman98}.

\bibitem{Friedman98} \Name{Friedman J.R., Sarachik M.P.\And Ziolo R.} \Review{Phys. Rev. B} \Vol{58} \Year{1998} \Page{R14729}.

\bibitem{Sun98} \Name{Sun Z.M. {\it et al.}}
\Review{Inorganic Chemistry} \Vol{37} \Year{1998} \Page{4758}.

\bibitem{Aubin97} \Name{Aubin S. M. {\it et al.}}
\Review{Chemical Communications} \Year{1997} \Page{2239}.

\bibitem{remark1}  The minor species of Mn$_{12}$ acetate are present in all currently 
synthesised crystals. Since it has first been mentioned by Sessoli {\it et al.} \cite{Sessoli93}, several 
authors realised the presence of these minor species \cite{Sun98, Aubin97} but they have never been studied in 
detail. Our main results of a detailed study, which will be published elsewhere [\Name{Barra A.L. {\it et al.}}, to be published], are resumed as follows: Currently available Mn$_{12}$ acetate crystals contain 5 to 8 \% of minor species which in our hypothesis correspond to 
defective sites in the crystal lattices showing modification of the coordination octahedron 
of a manganese(III) for the formation of  hydrogen bonds with the disordered carboxylic 
acid molecules of solvation. We call Mn$_{12}$ac(1) the major species, 
Mn$_{12}$ac(2) the minor species in the bulk of the crystals. While the anisotropy easy 
axis of Mn$_{12}$ac(1) is aligned with the c crystallographic axis, we found four easy 
axis of Mn$_{12}$ac(2) related by the four-fold symmetry axis of the tetragonal space 
group and forming an angle of $10^{\circ}$ with the c-axis.  The tilting of the anisotropy easy axis 
is due to the breaking of the tetragonal symmetry of the Mn$_{12}$ac molecule. The 
anisotropy barrier of Mn$_{12}$ac(2) is reduced to about 45 K (as determined through 
ac-measurements [\Name{Novak M. {\it et al.}}, to be published] and EPR [\Name{Barra A.L. {\it et al.}}, to be 
published] in comparison to 65 K for Mn$_{12}$ac(1). We found that Mn$_{12}$ac(2) 
is dispersed in a diluted way over the whole crystal and not forming domains as 
suggested in ref. 
[\Name{Takeda K., Awaga K.\And Inabe T.} \Review{Phys. Rev. B.} \Vol{57} \Year{1998} \Page{R11062}]. 
The SQUID array magnetometer allowed also to evidence other fast relaxing species 
mainly located at the surface of the crystal with no particular orientation of the easy axis. 

\bibitem{Ohm98} \Name{Ohm T., Sangregorio C.\And Paulsen C.} \Review{Europhysics J. B} \Vol{6} \Year{1998} \Page{595}; 
\Name{Ohm T., Sangregorio C.\And Paulsen C.} \Review{J. Low Temp. Phys.} \Vol{113} \Year{1998} \Page{1141}; 

\bibitem{Wernsdorfer99} \Name{Wernsdorfer W., Ohm T., Sangregorio C., Sessoli R., Mailly D. \And Paulsen C.} \Review{cond-mat/9901290}.

\bibitem{Prok98} \Name{Prokof'ev N.V. \And Stamp P.C.E.} \Review{Phys. Rev. Lett.} \Vol{80} \Year{1998} \Page{5794}; 
\Name{Prokof'ev N.V. \And Stamp P.C.E.} \Review{J. Low Temp. Phys.} \Vol{104} \Year{1996} \Page{143}.

\bibitem{remark2} We achieved a well define magnetisation state by different methods: 
(i) cooling the sample from 5 K to 40 mK in an applied field of H = 0 or -1.4 T, yielding 
the demagnetised or saturated magnetisation state of the entire crystal, respectively; (ii) 
first cooling the sample from 5 K to 40 mK in H = 0, then applying a field of H = -1.4 T 
for about 10 s, yielding a state where the major species Mn$_{12}$ac(1) is completely 
demagnetised whereas the minor species Mn$_{12}$ac(2) is saturated \cite{remark1}; 
(iii) one can also quench the sample from 5 K to 40 mK in a small field of few 10 $-$ 2 T. When the quench is fast (few seconds), the sample's magnetisation does 
not have time to 
relax, either by thermal or quantum transitions. This procedure yields a frozen thermal 
equilibrium distribution.

\bibitem{remark3} The 'hole digging' method is in analogy to the 'hole burning' method, 
currently used in high density optical storage 
[\Name{Hasan Z.,{\it et al.}} \Review{Applied Physics Letters} \Vol{72} \Year{1998} \Page{2373}] 
and single molecule spectroscopy 
[\Name{Xie X. S.\And Trautman J. K.} \Review{Annual Review of Physical Chemistry} \Vol{49} \Year{1998} \Page{441}]. 
The main difference is due to the fact that the hole is not only due to depleting of molecules in resonance at $H_{dig}$ but also to shifting other molecules away from resonance due to the dipolar coupling. 

\bibitem{Wernsdorfer96} \Name{W. Wernsdorfer} \Review{Ph.D. thesis} Joseph Fourier University, Grenoble, \Year{1996}.

\bibitem{remark4} For these measurements, we used an array of micro-SQUID 
\cite{Wernsdorfer96} with a very high sensitivity. We could investigate singles crystals 
10 to 100 $\mu$m in size. The magnetometer is working in the temperature range 
between 35 mK and 6 K. Field sweeping rates can be as high as 1 T/s with a maximum 
field of 1.4 T. The time resolution is about 1 ms allowing short-time measurements.

\bibitem{remark5} The relaxation curves are better fit to a power law $t^\alpha$ with $0.3 < \alpha < 0.5$. This is irrelevant to the discussion of the letter because it hardly effects the hole line widths which is the main point of this letter.

\bibitem{Hartmann96} \Name{Hartmann-Boutron F., Politi P.\And Villain J.} \Review{Int. J. Mod. Phys.} \Vol{10} \Year{1996} \Page{2577}.
\bibitem{Zheng96} \Name{Zheng M. \And Dismukes G. C.} \Review{Inorg. Chem.} \Vol{35} \Year{1996} \Page{3307}.

\bibitem{Leuenberger99} \Name{Leuenberger M.N. \And Loss D.} \Review{cont-mat/9810156}.

\bibitem{Fort98} \Name{Fort A., Rettori A., Villain J., Sessoli R. \And Gatteschi D.} \Review{Phys. Rev. Lett.} \Vol{80} \Year{1998} \Page{612}.

\bibitem{remark6} This reasoning supposes that the inhomogeneous hyperfine 
broadening of about 12 mT is temperature independed in the region between 
0.04 and 3 K.

\end{thebibliography}
\end{document}